\documentclass[aip,rsi,reprint]{revtex4-1}
\usepackage{amsmath}
\usepackage{graphicx}
\usepackage{epstopdf}
\usepackage{hyperref}
\usepackage{natbib}
\usepackage{color,soul}
\usepackage{ragged2e,array,booktabs}

\let\oldhat\hat
\renewcommand{\vec}[1]{\mathbf{#1}}
\renewcommand{\hat}[1]{\oldhat{\mathbf{#1}}}

\begin{document}


\title{Dielectric Resonator Method For Determining Gap Symmetry Of Superconductors Through Anisotropic Nonlinear Meissner Effect}


\author{Seokjin Bae}
\thanks{Author to whom correspondence should be: sjbae@umd.edu}
\affiliation{Center for Nanophysics and Advanced Materials, Department of Physics, University of Maryland, College Park, MD, 20742-4111, USA}

\author{Yuewen Tan}
\affiliation{Center for Nanophysics and Advanced Materials, Department of Physics, University of Maryland, College Park, MD, 20742-4111, USA}

\author{Alexander P. Zhuravel}
\affiliation{B. Verkin Institute for Low Temperature Physics and Engineering, National Academy of Sciences of Ukraine, UA-61103 Kharkov, Ukraine}

\author{Lingchao Zhang}
\affiliation{Department of Physics, National University of Singapore 117551, Singapore}
\affiliation{NUSNNI-Nanocore, National University of Singapore 117411, Singapore}

\author{Shengwei Zeng}
\affiliation{Department of Physics, National University of Singapore 117551, Singapore}
\affiliation{NUSNNI-Nanocore, National University of Singapore 117411, Singapore}

\author{Yong Liu}
\affiliation{Ames Laboratory, Ames, IA 50011, USA}

\author{Thomas A. Lograsso}
\affiliation{Ames Laboratory, Ames, IA 50011, USA}

\author{Ariando}
\affiliation{Department of Physics, National University of Singapore 117551, Singapore}
\affiliation{NUSNNI-Nanocore, National University of Singapore 117411, Singapore}

\author{T. Venkatesan}
\affiliation{Department of Physics, National University of Singapore 117551, Singapore}
\affiliation{NUSNNI-Nanocore, National University of Singapore 117411, Singapore}

\author{Steven M. Anlage}
\affiliation{Center for Nanophysics and Advanced Materials, Department of Physics, University of Maryland, College Park, MD, 20742-4111, USA}



\date{\today}

\begin{abstract}
  We present a new measurement method which can be used to image gap nodal structure of superconductors whose pairing symmetry is under debate. This technique utilizes a high quality factor microwave resonance involving the sample of interest. While supporting a circularly symmetric standing wave current pattern, the sample is perturbed by a scanned laser beam, creating a photoresponse that was previously shown to reveal the superconducting gap anisotropy. Simulation and the measurement of the photoresponse of an unpatterned Nb film show less than 8\% anisotropy, as expected for a superconductor with nearly isotropic energy gap along with expected systematic uncertainty. On the other hand, measurement of a YBa$_2$Cu$_3$O$_{7-\delta}$ thin film shows a clear 4-fold symmetric image with $\mathtt{\sim}12.5$\% anisotropy, indicating the well-known 4-fold symmetric $d_{x^2-y^2}$ gap nodal structure in the $ab$-plane. The deduced gap nodal structure can be further cross-checked by low temperature surface impedance data, which is simultaneously measured. The important advantage of the presented method over the previous spiral resonator method is that it does not require a complicated lithographic patterning process which limits one from testing various kinds of materials due to photoresponse arising from patterning defects. This advantage of the presented technique, and the ability to measure unpatterned samples such as planar thin films and single crystals, enables one to survey the pairing symmetry of a wide variety of unconventional superconductors.
\end{abstract}


\maketitle

\section{Introduction}
\par Among the several quantities that characterize superconductors, the superconducting gap function in momentum space is one of the most important parameters which governs the phenomenon. The symmetry of this gap function is directly related to the symmetry of the wavefunction of the superconducting electron pairs.\cite{Sigrist1991RMP} Many unconventional superconductors have nodes in their superconducting gap function $\Delta\left(\vec{k}\right)$ which are robust and arise from the symmetry of the pairing wavefunction. Thus, determining this gap nodal structure can give a significant clue about the pairing mechanism for the material.

\par  Because of this importance, there have been numerous measurement methods developed to map out the gap structure on the Fermi surface.\cite{CCTsuei2000RMP} For example, Raman scattering,\cite{Devereaux1994PRL} angle-resolved photoemission spectroscopy (ARPES),\cite{ZXShen1993PRL} angle-resolved specific heat measurement (ARSH),\cite{Aubin1997PRL,Revaz1998PRL,Matsuda2006JPhys} and superconducting quantum interference device (SQUID) interferometry\cite{DJVHarlingen1995RMP} are commonly used. However, each method has advantages and disadvantages (see Table \ref{table1}) and it is best to use multiple methods to develop a consistent and complete picture of gap symmetry.  ARPES and SQUID interferometry are sensitive to near-surface properties so they require very clean surfaces or high-quality tunnel barriers. Many materials present surfaces that are not characteristic of the bulk, or do not make high quality tunnel junctions, and this drawback limits them to study a relatively small number of materials. Meanwhile, ARSH and to some extent Raman scattering measure bulk response so they are free from the near-surface sensitive issue. However, they investigate quasiparticle response whose anisotropy is generally weaker than that of the superfluid response. In addition, ARSH depends on the presence of magnetic vortices. Also, to interpret the data from these methods, detailed information about the Fermi surface is required. To augment and partially overcome the limitations of these techniques, a new gap nodal spectroscopy method using the anisotropic nonlinear Meissner effect (aNLME) was proposed theoretically by Yip, Sauls, and Xu\cite{Yip1992PRL, DXu1995PRB}, Dahm and Scalapino\cite{Dahm1996APL, Dahm1997JAP}, and manifested in the experiment of Zhuravel et al.\cite{Benz2001PhysicaC,Oates2004PRL,Zhuravel2013PRL,Zhuravel2018PRB} This new gap nodal spectroscopy using the aNLME gives an image of the gap nodal structure from both the bulk superfluid response and the surface Andreev bound state response, for appropriate surfaces.\cite{Zhuravel2013PRL,Zhuravel2018PRB}

\begin{widetext}
\begingroup
\squeezetable
\begin{table}
\centering
\begin{tabular}{ >{\RaggedRight}p{1.1cm} @{\qquad} >{\RaggedRight}p{6.6cm} @{\qquad} >{\RaggedRight}p{7.8cm}}
\hline
\hline
Technique & Advantages & Disadvantages \\
\hline
ARPES\cite{ZXShen1993PRL} & Directly image band structure        
        and gap $\Delta(\vec{k})$ & Requires very pristine surfaces, finite energy resolution \\
\hline
SQUID\cite{DJVHarlingen1995RMP} & Sensitive to the sign change of the gap $\Delta(\vec{k})$ & Requires high-quality tunnel junctions \\
\hline
ARSH\cite{Revaz1998PRL} & Relatively simple thermodynamic measurement & Depends on the presence of magnetic vortices  \\
& & Interpretation is dependent on knowledge of Fermi surface details \\
\hline
Raman\cite{Devereaux1994PRL} & Able to choose specific symmetries under test by choosing polarization orientations  & Requires detailed theoretical calculations of response functions for each polarization orientation to interpret data \\
\hline
aNLME\cite{Benz2001PhysicaC,Oates2004PRL,Zhuravel2013PRL,Zhuravel2018PRB} & Directly image gap nodal structure in real-space, not sensitive to near-surface quality & Requires high-Q resonance with circulating currents over a single-domain sample.\\
\hline
\hline
\end{tabular}
\caption{\label{table1} Summary of some advantages and disadvantages of several representative superconducting gap spectroscopy techniques (not an exhuastive list).}
\end{table}
\endgroup
\end{widetext}

\par The principle of how the NLME brings out the gap nodal structure is as follows. When external magnetic field is applied to a superconductor, it generates a supercurrent to expel the fields, a hallmark of superconductivity known as the Meissner effect. The kinetic energy invested in this screening current diminishes the difference in free energy between the superconducting and normal states. In terms of Ginzburg-Landau theory, this results in a decrease in the superconducting order parameter, accompanied by a decrease in the superfluid density $n_s(T,\vec{j})$. As a result, the magnetic penetration depth $\lambda(T,\vec{j})\sim 1/\sqrt{n_s(T,\vec{j})}$ increases correspondingly. This nonlinear effect can be parametrized in terms of the current density $\vec{j}$. In the regime where the screening current density is small compared to the zero temperature critical current density ($j/j_{c}(0) \ll 1$), the nonlinearity can be expressed as follows\cite{Dahm1996APL}
\begin{gather} \label{superfluid density in terms of Meissner coefficient}
n_s(T,\vec{j}) \cong n_s(T)\left( 1-b_\Theta(T)\left( \frac{\vec{j}}{j_c(0)} \right)^2 \right) \\
\lambda^2(T,\vec{j}) \cong \lambda^2(T)\left( 1+b_\Theta(T)\left( \frac{\vec{j}}{j_{c}(0)} \right)^2 \right),
\end{gather} 
where $b_\Theta(T)$ is the nonlinear Meissner coefficient, $\Theta$ represents the direction of the superfluid velocity relative to a reference direction of the gap in k-space, and $\vec{j}$ is the vector current density. Note that the anisotropy of $b_\Theta(T)$ in k-space is directly determined by the nodal structure of the gap function $\Delta(\vec{k})$.\cite{Dahm1996APL,Dahm1997JAP} For example, if the gap function has four nodes and anti-nodes, which is the case of a $d_{x^2-y^2}$-wave superconductor, $b_\Theta(T)$ is also 4-fold symmetric.\cite{Dahm1996APL} Vice versa, if one can image the anisotropy of the $b_\Theta(T)$ in k-space, one can deduce the gap nodal structure.

\section{Principle of the experiment}\label{ExpPrinciple}
\begin{figure}
		\includegraphics[width=\columnwidth]{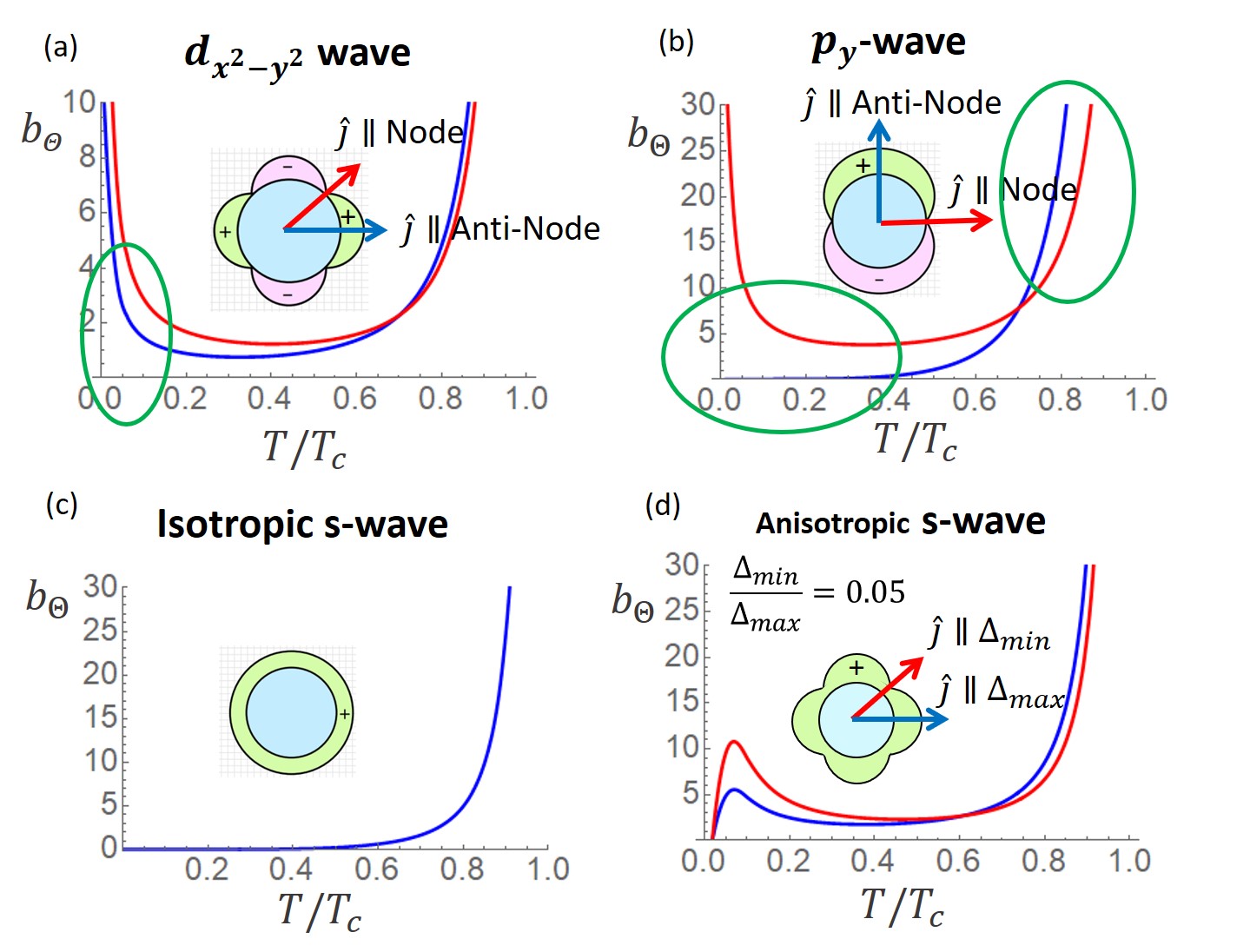}
		\caption{\label{fig:bTheta} Temperature dependence of the nonlinear Meissner coefficient $b_\Theta(T)$ when the direction of the current is parallel to the gap nodal direction (red) and anti-nodal direction (blue) for the case of (a) $d_{x^2-y^2}$-wave superconductor, (b) $p_y$-wave superconductor, (c) isotropic $s$-wave superconductor, and (d) anisotropic $s$-wave superconductor. For the $s$-wave case, since there is no gap nodal direction, only one curve is plotted. For the case of the anisotropic $s$-wave case, the red line represents the gap minimum direction and the blue line represents the gap maximum direction. The green ovals in (a), (b) denote the temperature regime where the anisotropy in $b_\Theta$ is large.}
\end{figure}

\par Experimentally, to image the anisotropy in the $b_\Theta$, a thermal perturbation method is used. For the case of gap nodal superconductors such as the $d$- or $p$-wave cases, $b_\Theta(T)$ shows a large anisotropic temperature dependence at low temperature $T/T_c<0.2$ between situations when the direction of the current is parallel to the gap nodal and anti-nodal direction (Fig.\ref{fig:bTheta}(a),(b)). On the other hand, for the case of an $s$-wave or slightly anisotropic $s$-wave superconductor, the temperature dependence of $b_\Theta(T)$ is isotropic and weak for $T/T_c<0.2$ (Fig.\ref{fig:bTheta}(c)). However, a strongly anisotropic $s$-wave superconductor will reveal anisotropic $b_\Theta(T)$ at low temperature (Fig.\ref{fig:bTheta}(d)). Under a local thermal perturbation which modulates the temperature at a point on the sample, the temperature derivative of $b_\Theta$ governs the sample nonlinear response. As a result of heating, the modulation in the local superfluid density $\delta n_s(T,\vec{j})$ inherits the anisotropy in $db_\Theta/dT$,\cite{Dahm1997JAP,Zhuravel2013PRL} hence  creating anisotropy in the electromagnetic response of the superconductor. 

\begin{figure}
		\includegraphics[width=\columnwidth]{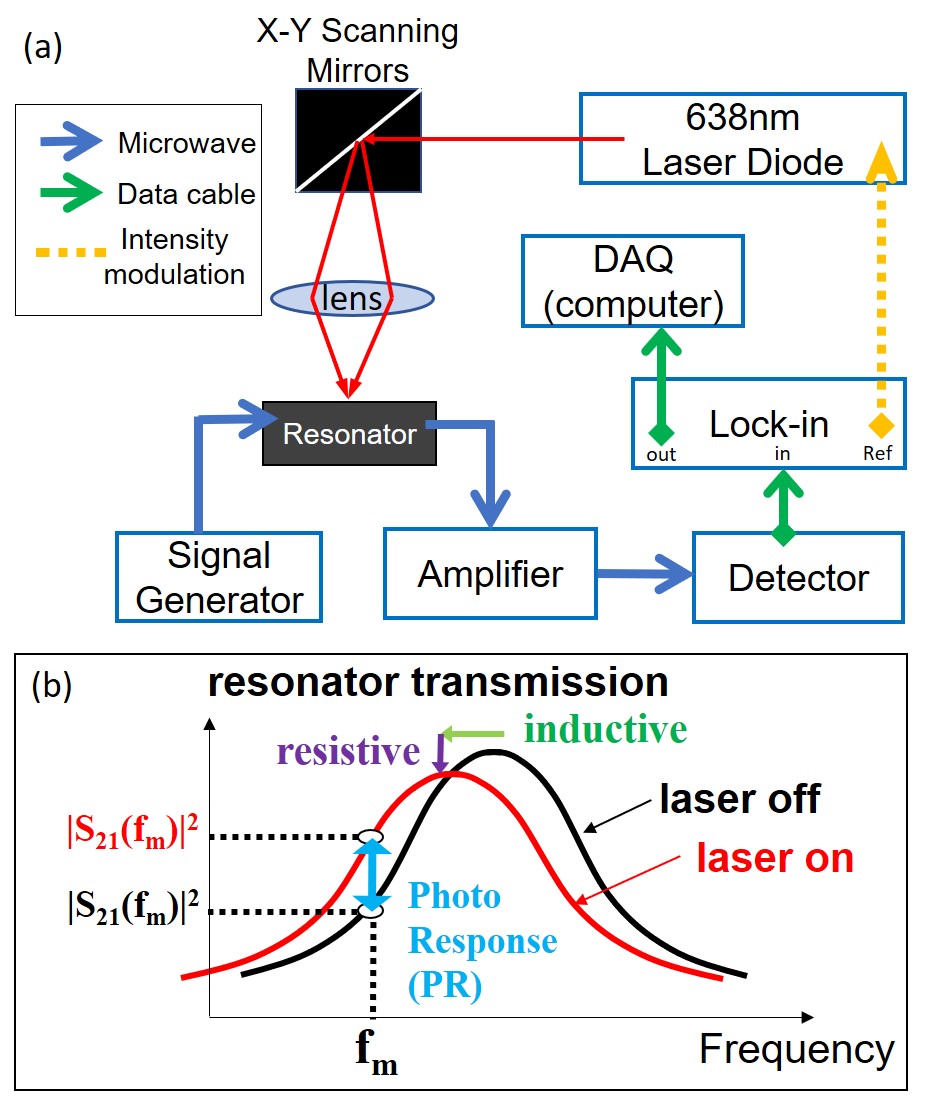}
		\caption{\label{fig:blockdiagram} (a) Block diagram and procedure of LSM-PR measurement. The path for each type of signal is described in the box inset. (b) Schematic resonator microwave transmission when the laser is on (red) and off (black). Here, $f_m$, the optimal measurement frequency, is chosen where PR shows its maximum as a function of frequency. The contribution to PR from the shift in the resonance frequency $f_0$ is denoted as the inductive PR ($PR_X$) and the contribution to PR from the change in the quality factor $Q$ is denoted as the resistive PR ($PR_R$).}
\end{figure}

\par The anisotropy of the electromagnetic response can be investigated by a laser scanning microscope (LSM). As seen from Fig.\ref{fig:blockdiagram}(a), a microwave signal is injected into a resonator containing a superconducting sample. Below $T_c$ and for specific resonant modes, a steady state circulating microwave current is induced on the surface of the sample (discussed in detail in Sec.\ref{DRDesignSimul}). Then, a focused scanning laser beam illuminates a small part of the sample and the laser intensity is modulated periodically in time. The illumination induces a local temperature modulation $\delta T$ that is periodic in time. Since the local superfluid density $n_s$ depends on the local temperature, the temperature change brings about a change in the local penetration depth $\lambda$ (and hence the inductance) and local surface resistance $R_s$ (and hence the microwave loss) of the sample in a manner that depends on the local direction of the microwave current. The change in the inductance shifts the resonant frequency $f_0$ and the change in the loss affects the quality factor $Q$ of the resonance, which results in a change in the microwave transmission at a fixed measurement frequency $f_m$\cite{KaiserAPL1998,Zhuravel2006APL} (see Fig.\ref{fig:blockdiagram}(b)). This modulation in transmission induced by the laser illumination is called photoresponse (PR) and is measured by a narrow-band phase-sensitive method using a lock-in amplifier. The contribution to PR from the inductance change is denoted as inductive PR ($PR_X$) and the contribution to PR from the resistance change is denoted as resistive PR ($PR_R$). These two contributions can be modeled as,\cite{Culbertson1998JAP,Zhuravel2006APL}
\begin{eqnarray} 
PR_X \sim |\vec{j}|^2\delta\lambda \label{PR_X} \\
PR_R \sim |\vec{j}|^2\delta R_s \label{PR_R}
\end{eqnarray}
where $\delta\lambda$, $\delta R_s$ are the change in the local penetration depth and surface resistance due to the laser illumination (which depend on the direction of $\vec{j}$), and $|\vec{j}|^2$ is the local value of current density squared.

\par The laser beam is then systematically scanned over the sample and the PR is measured at each dwell location of the beam. By imaging the photoresponse arising from the circulating current, which surveys the entire in-plane Fermi surface, one can map out the anisotropy of the electromagnetic response. This anisotropy is governed by that of the modulation in superfluid density $\delta n_s(T)$ and hence that of the temperature derivative of the nonlinear Meissner coefficient $db_\Theta/dT$,\cite{Dahm1997JAP,Zhuravel2013PRL} which ultimately arises from the anisotropy of $\Delta(\vec{k})$. Therefore, it is essential to ensure that one induces a uniform circulating current on the sample, such that it uniformly samples the entire in-plane Fermi surface. In that context, the design of the resonator becomes crucial for the validity of this measurement technique.

\begin{figure}
		\includegraphics[width=\columnwidth]{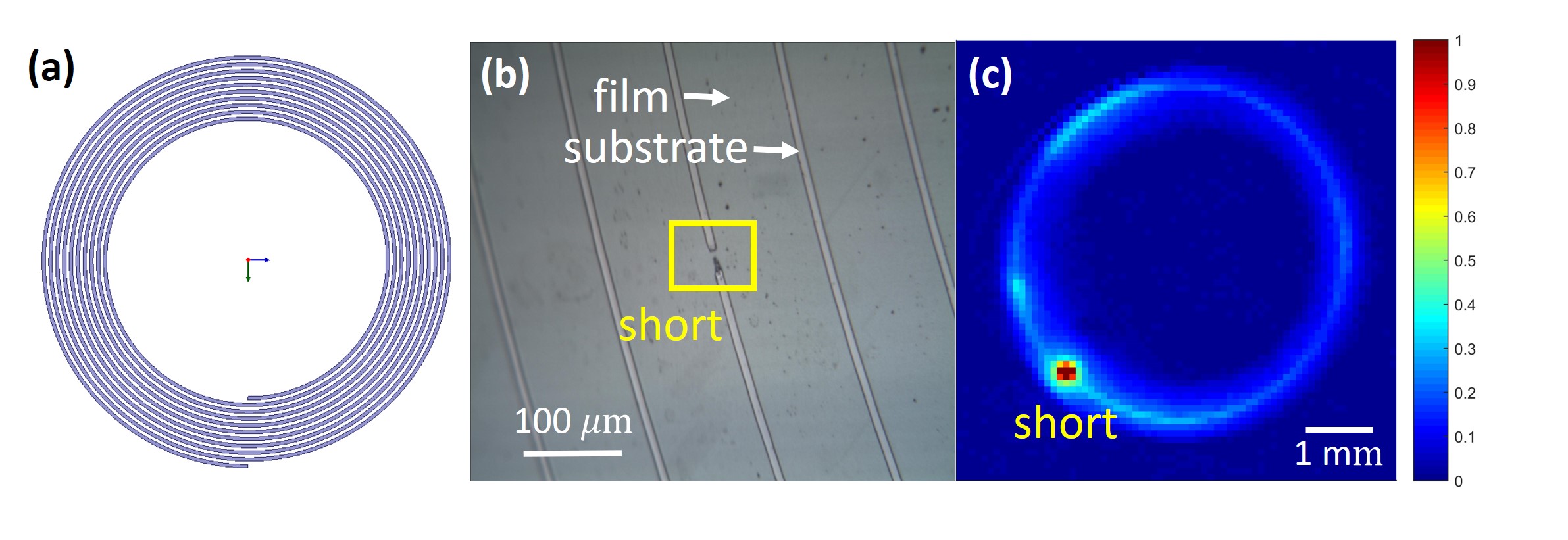}
		\caption{\label{fig:spiralPR} (a) Schematic of the spiral structure (Not to scale). The spiral has inner diameter of 4 mm, outer diameter of 6 mm, $8\sim40$ turns typically, and thickness of 300 nm. (b) Example real-space image of the defect on a Pr$_{1.85}$Ce$_{0.15}$CuO$_4$ (PCCO) spiral sample. The line width of the spiral strip is 100 $\mu$m, the spacing between the strips is 10 $\mu$m, and the number of turns is 8.5. Note the presence of a short circuit between strips in the patterned spiral. (c) A defect dominated PR image obtained from the PCCO sample at its fundamental resonance mode ($\approx$ 383 MHz) at 5.48 K. Note the correspondence of the defect (short) in (b) with the PR hotspot due to the short in (c). This hotspot overwhelms PR from other parts of the spiral and obscures observation of the gap symmetry from the PR images. }
\end{figure}

\par The first-generation resonator design which satisfies the above requirements is the multi-turn spiral resonator (Fig.\ref{fig:spiralPR}(a)).\cite{Kurter2010APL,Kurter2011IEEE,Zhuravel2012PRB,Ghamsari2013APL} The spiral resonator is prepared by patterning a thin-film sample into an Archimedean spiral shape with a photo-lithographic procedure. At its resonant frequencies, this spiral geometry has the advantage that it produces a uniform tangential standing wave current. Another advantage is that it is a self-resonant structure which produces a strong response when perturbed by the laser. Thus, one can obtain a clear photoresponse image which reveals the gap structure throughout the in-plane Fermi surface of a structurally coherent superconductor, as proven with c-axis oriented YBa$_2$Cu$_3$O$_{7-\delta}$ (YBCO) films.\cite{Zhuravel2013PRL,Zhuravel2018PRB} However, despite the advantages, it has one significant drawback. Patterning a spiral shape on a thin film sample requires an elaborate and potentially destructive lithographic process. During this process, it is not only easy to degrade the superconducting properties ($T_c$, for example) of the film but also easy to create defects (holes, cuts, or shorts etc) on the spiral [Fig.\ref{fig:spiralPR}(b)]. These defects induce PR hotspots which dominate the entire PR image and make it difficult to deduce the gap symmetry from the image. This drawback is demonstrated in Fig.\ref{fig:spiralPR}(c) and in other papers that discuss this issue\cite{Kurter2011PRB,Zhuravel2012PRB,Ciovati2012RSI}. In fact, many of the newly emerged unconventional superconductors of interest are very vulnerable to this defect issue during the lithographic process due to their sensitivity to solvents, water, and the ambient atmosphere. Also of concern, some materials grow as an incoherent, multi-domain disordered structure when they are prepared in a thin film form, and this prevents their examination with this technique because a single-domain sample of the size of the resonator is required. Therefore, a new resonator design is needed which can examine unpatterned thin films and eliminate the lithography-induced defect hotspot issue. In addition, a resonator design which can examine single crystals is most desirable since some unconventional superconductors are sensitive to non-magnetic disorder and cannot be prepared in a thin film form.

\section{Dielectric Resonator Method}
\subsection{Design and Simulation}\label{DRDesignSimul}

\begin{figure}
		\includegraphics[width=\columnwidth]{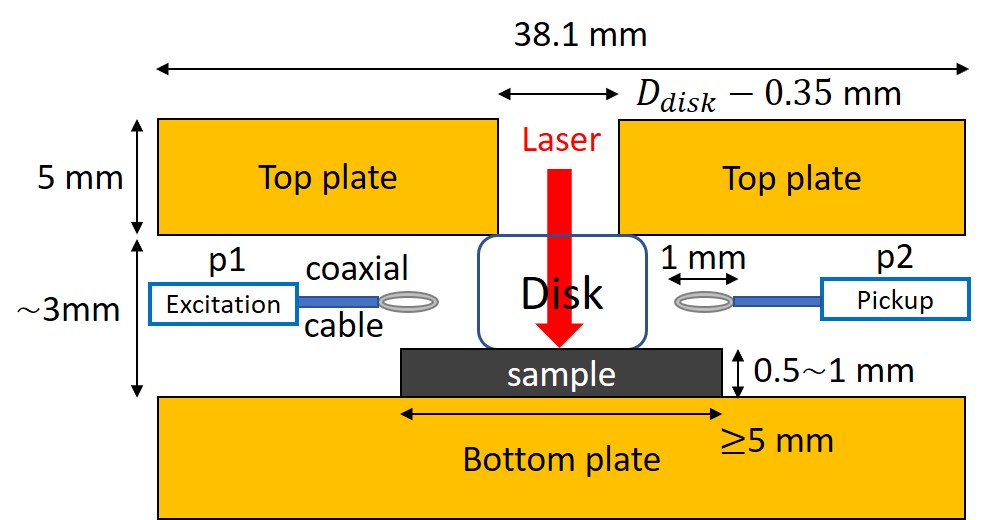}
		\caption{\label{fig:DR} Schematic cross-section diagram of the dielectric resonator setup with a sample (not to scale). The transparent dielectric disk is sandwiched between the top plate and the sample. The diameter of the dielectric disk $D_{disk}$ is either 6.35 mm (sapphire disk) or 3 mm (rutile disk). The diameter of the aperture in the top plate is 0.35 mm smaller than $D_{disk}$. The resonator is coupled to microwave coaxial cables through magnetic loops. }
\end{figure}

\par To satisfy the above needs to measure unpatterned films and single crystals while maintaining the requirements for the gap spectroscope via PR measurement, the dielectric resonator (DR) design with an aperture (see Fig.\ref{fig:DR}) is adopted. This DR is a modified version of the Hakki-Coleman type resonator.\citep{HakkiColeman1960,Mazierska1998IEEE} 
It consists of a top and bottom metallic plate (Cu or Nb) which confine the microwave fields inside the resonator like a cavity. A cylindrical disk with high dielectric constant, which is placed on top of a superconducting sample, creates a resonance that induces strong microwave currents on the sample. The resonant frequency $f_0$ is determined mainly by the dimension and the dielectric constant of the disk. The un-patterned sample is placed in contact with one face of the disk to modify the resonant properties of the DR.

\par The dielectric material must satisfy three requirements. First is a high dielectric constant $\epsilon_r$ that is isotropic in the plane of the sample to concentrate the microwave fields to a small part of the sample, and the second is a low loss tangent ($\tan\delta$) at cryogenic temperatures to enable a high quality factor for a microwave resonance. The third requirement is to be transparent at the wavelength of the laser used for thermal perturbation. Sapphire and rutile are the best choices satisfying these requirements. Sapphire has high dielectric constant ($\epsilon_{a,b,c}\sim 10$ where $a$, $b$ are the in-plane crystallographic axes and $c$ is the out-of-plane axis) and very low loss tangent ($\mathtt{\sim}10^{-10}$) at temperatures below 10 K.\cite{Shelby1980JPCS,Tobar1998IEEE,Huttema2006RSI} Rutile has even higher dielectric constant ($\epsilon_{c} > 250$, $\epsilon_{a,b} > 120$) and still low loss tangent ($\mathtt{\sim}10^{-8}$) at temperatures below 10 K.\cite{Sabisky1962JAP,Tobar1998JAP} Due to the difference in the dielectric constant, sapphire is suitable for a large or homogeneous sample ($\mathtt{\sim}$10$\times$10 mm$^2$) and rutile is suitable for a small sample ($\mathtt{\sim}$5$\times$5 mm$^2$). Both materials are transparent to visible light.

\begin{figure}
		\includegraphics[width=\columnwidth]{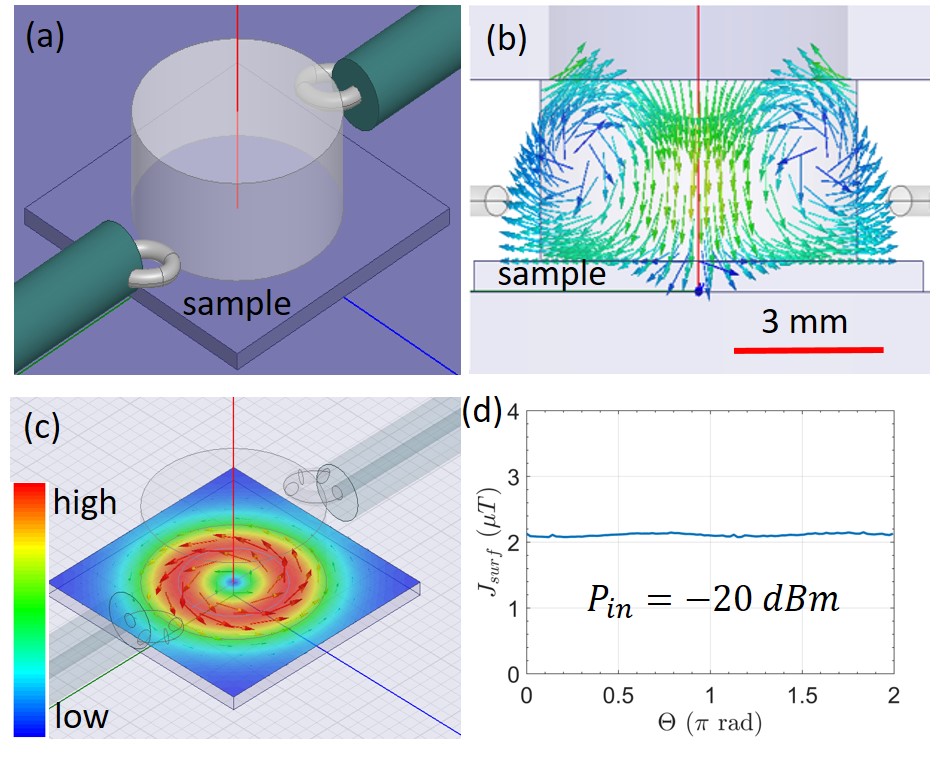}
		\caption{\label{fig:simulation} (a) Numerical simulation (HFSS) setup geometry of DR with a sapphire disk (top plate is hidden from view). Here, a perfect electric conductor boundary condition is imposed on the sample surface to represent the response of a superconducting sample. (b) Microwave magnetic field density plot inside the dielectric disk in the TE$_{011}$ resonance mode ($f_0 \sim 20$ GHz). (c) Surface current density plot on the sample and (d) its angular dependence for an input RF power $P_{in}=-20$ dBm. The induced current density angular ($\Theta$) dependence is nearly constant.}
\end{figure}

\par Among the many resonant modes generated by the cylindrical dielectric resonator, the TE$_{011}$ mode\cite{Kajfez1984IEEE} provides the desired field configuration on the sample surface. The microwave magnetic field in this mode has a toroidal shape as shown from an HFSS simulation (Fig.\ref{fig:simulation}(b)). In response to this field, the superconducting sample generates a current which circles around the axial line of the cylindrical disk to screen out the field (Fig.\ref{fig:simulation}(c)). Fig.\ref{fig:simulation}(d) shows that this rf current distribution is uniform in its angular distribution.

\par Once this resonant circulating current is induced, a focused modulated laser beam (638 nm) illuminates a point of the sample through the aperture in the top plate of the DR and the optically transparent dielectric disk. Simulations show that opening the aperture in the top plate does not significantly reduce the quality factor of the resonator for frequencies below the cutoff of the empty cylindrical waveguide. The beam is scanned across the sample and PR is collected from all directions of the standing wave current. Under appropriate circumstances, the resulting PR image reflects the anisotropy of $db_\Theta/dT$, which can be related to the gap nodal structure of the sample.

\par To verify the validity of the presented resonator design as a new gap spectroscope, one must make sure that there is no resonator-geometry induced anisotropy in the circulating current distribution. As seen from the expression to estimate the inductive and resistive PR in Eqs.(\ref{PR_X}),(\ref{PR_R}), the anisotropy of PR can arise from both intrinsic ($\delta\lambda$, $\delta R_s$) and extrinsic ($|\vec{j}|^2$) origins. The anisotropy in $\delta\lambda$ and $\delta R_s$ is introduced from the gap $\Delta(\vec{k})$ which is encoded in $b_\Theta$. However, there can also be anisotropy in $|\vec{j}|^2$ introduced by the possible asymmetries of the geometry of the resonator. To fully claim that measured PR anisotropy is equivalent to that of the gap function, it should be proven that the geometric anisotropy of the DR is negligible compared to the expected PR anisotropy for the case of a nodal gap superconductor. In the following, we use a combination of simulation and experiment to prove the claim.   

\subsection{Estimation of Systematic Uncertainty in Anisotropy of PR due to Angular Distribution of the Current Density}\label{CurrentDensityAnisotropy}

\begin{figure}
		\includegraphics[width=\columnwidth]{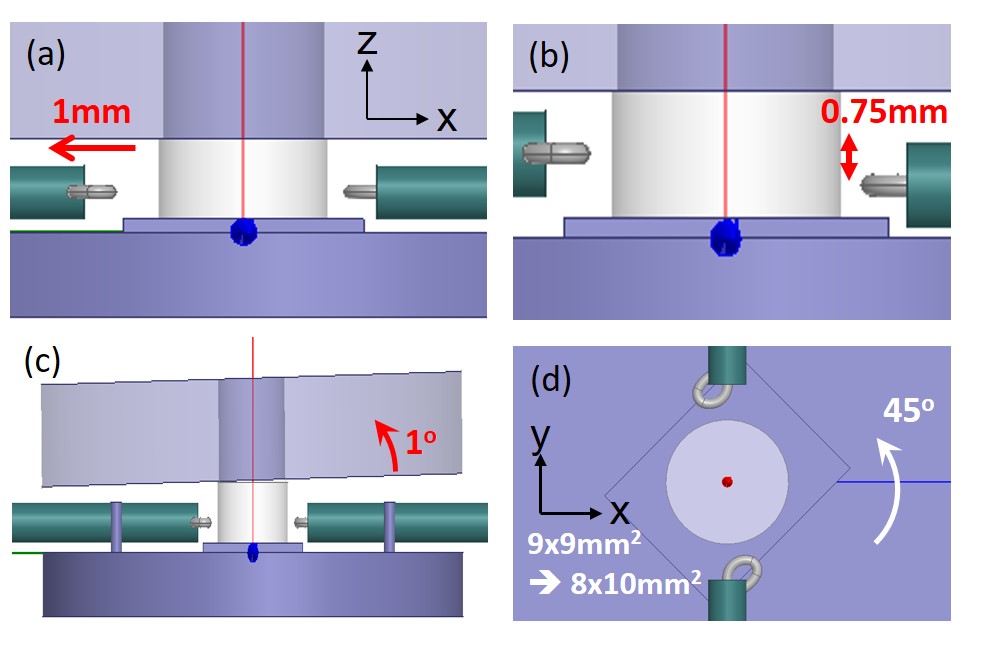}
		\caption{\label{fig:robustness} Simulated setup asymmetries to test the robustness of the uniformity of the angular dependence of the current density. (a) A shift in the loop position by 1 mm in the x-direction, (b) by 0.75 mm in the z-direction, (c) a tilted top plate by an angle of 1$^\circ$, (d) rectangular shaped sample with 45 degrees counterclockwise rotation.}
\end{figure}

\par To validate the absence of geometric anisotropy, a current density $\vec{j}$ on an isotropic sample (assumed perfect conductor) is simulated with HFSS. We want to examine the degree to which $|\vec{j}_\Theta|=|\vec{j}(r_m,\Theta)|$ is a uniform function of angle $\Theta$, where $r_m$ is the radius which gives the largest $|\vec{j}(r,\Theta)|$ as a function of radius $r$ (The red circular region in Fig.\ref{fig:simulation}(c)). In the simulation, Cu is used for the top and bottom plates, and a sapphire disk whose $c$-axis is aligned with the cylindrical axis of the disk is used for the dielectric material, with a diameter of 6.35 mm and a height of 3 mm. With these dimensions, the TE$_{011}$ mode occurs at $\mathtt{\sim}20$ GHz which is in a typical operating frequency for microwave transmission lines and devices. The lateral dimension of the sample is 9$\times$9 mm$^2$ to fully screen the field from the DR. As seen from Fig.\ref{fig:simulation}(c),(d), the magnitude of the current density along the circle is uniform, and its angular dependence shows only a 2.8\% anisotropy ratio, which is defined as $(|\vec{j}_\Theta|^2_{max} - |\vec{j}_\Theta|^2_{min})/|\vec{j}_\Theta|^2_{min}$ around the circle. This small anisotropy occurs due to an effect of the coupling loops to the microwave field distribution which is small but breaks cylindrical symmetry. To further ensure the robustness of the uniformity of the current density, various kinds of possible geometric asymmetries are imposed on the resonator. As seen from Fig.\ref{fig:robustness}, we consider displacements in one of the loop positions (shift in the x, z-direction), a tilt of the top plate, and a rectangular shaped sample with a 45-degree rotation. For each of these asymmetries, the simulations show the anisotropy ratio of 6.6\%, 5.2\%, 5.5\%, and 7.3\%, respectively. This result establishes the typical scale for geometric anisotropy in the presented dielectric resonator design in a simulational aspect. If one observes larger anisotropy in the PR image than this geometric anisotropy scale, the observed anisotropy in PR can be due to the gap nodal structure of the sample. In Sec.\ref{PR result}, the degree of geometric anisotropy will be established again through experimental tests.

\section{Measurement setup}
\subsection{Cryogenics and microwave measurement setup}
\par For the measurement of PR in various temperature ranges below $T_c$, the DR is mounted on one of two types of optical cryostat. One is a BlueFors XLD-400 cryostat which has a temperature range of 40 mK $\sim$ 10 K at the sample stage, and the other is an ARS cryostat with $6\sim300$ K sample temperature range.\cite{Remillard2014SST,Ghamsari2016SciRep} Each cryostat has an optical window which allows scanned laser illumination for the thermal perturbation. Here, we focus on the LSM in the BlueFors cryostat. As seen from the block diagram in Fig.\ref{fig:blockdiagram}, an Omicron LuxX+ 638 nm laser source is intensity modulated up to 1.5 MHz by an SRS865 lock-in amplifier. The input microwave signal is generated by an Agilent E8257C signal generator. The transmitted signal is amplified by an LNF-LNC6-20C cryogenic amplifier and measured by an HP 8473C microwave detector. This signal is fed to the lock-in to measure the PR.

\begin{widetext}

\begin{figure}
		\includegraphics[width=1\columnwidth]{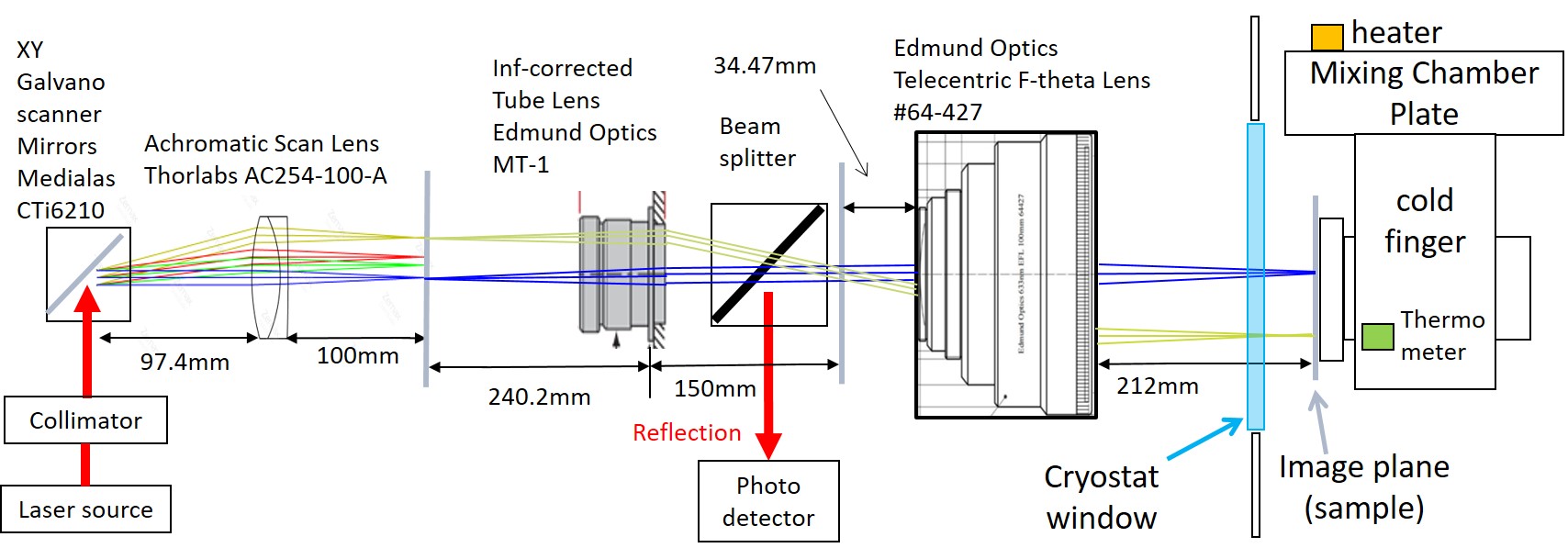}
		\caption{\label{fig:4fscanning} Schematic view of optical path of the laser beam in the 4f scanning system. The grey vertical lines are focal planes. Not to scale. }
\end{figure}
\end{widetext}

\subsection{4f laser scanning system}
\par To obtain a 2D PR image from the circulating current distribution of the sample surface, a 4f laser scanning system is used. As depicted in Fig.\ref{fig:4fscanning}, the 638 nm laser beam from the laser source is collimated to a 2 mm diameter 1/e$^2$ Gaussian beam, and incident on two closely-spaced galvanometric mirrors (XY scanner). The deflection angle of the mirrors, which ranges through $\pm 10^\circ$, is controlled by voltage applied to their motors. The deflected scanning beam first passes though an achromatic scan lens, and then is directed into an infinity corrected tube lens (ITL) with long focal length (200 mm) so that the transmitted beam is collimated again with almost two times smaller divergence. The collimated beam is then incident on the entrance pupil of the telecentric f-theta lens. The telecentric f-theta lens directs the beams with different incident angle into orthogonally focused XY scanning beams with parallel translation paths in the image plane. The position of the spot (parallel translation of the beam) on the image plane is linearly proportional to the scan angle. This ensures the beams with different translated path share a flat (not curved) image plane, yielding a low distortion of the beam spot size. This ensures the beam intensity on the sample surface remains constant while scanning. The cryostat has optical windows for visible light which let the beams from the f-theta lens illuminate the sample. The working distance (212 mm) of the f-theta lens is long enough to have an image plane on the sample holder inside the cryostat.

\par After the beam applies a thermal perturbation to the sample to generate PR, the reflected beam pass through the telecentric f-theta lens which makes the beam follow the original incident path until it encounters a beam splitter (Edmund Optics \#54-823) located in between the ITL and f-theta lens. The reflected beam is then guided to a photo-detector (Edmund Optics \#53-373) which is connected to a second lock-in amplifier that is referenced to the light modulation. The obtained reflectivity image contains optical microscope information of the sample surface so that one can align the PR image to a real-space image of the sample. In addition, it can relate any artifacts in the PR image arising from optical interference patterns (such as Newton rings) or mechanical defects like a scratch on the sample surface.

\section{Experimental results} \label{PR result}
\subsection{Photoresponse measurement with the dielectric resonator} \label{PR images}
\par The first validation required for the DR design to be utilized as a gap spectroscope is to experimentally prove that it imposes only marginal geometric anisotropy to the measured PR. To do this, the PR is measured and imaged from a Nb sample whose superconducting gap has small ($\lesssim9\%$)\cite{Macvicar1967} anisotropy. Since the anisotropy in $b_\Theta(T)$ is only $< 0.4$\% for this case at the measurement temperature ($\sim 8$ K), the anisotropy in the PR image from the Nb sample should arise mainly from the geometric anisotropy of the resonator.

\par The Nb thin film sample is grown on a silicon substrate by Ar sputtering with a thickness of 300 nm, $T_c$ = 9.25 K, and high $RRR=100$. With this sample mounted and with the same sapphire disk and Cu top plate described above, a high $Q$ ($>10^4$) resonance is obtained at around 20.6 GHz and 4 K. Because an s-wave superconductor is predicted to have stronger PR near $T_c$ (Fig.\ref{fig:bTheta}(c)),\cite{Dahm1996APL} PR is measured around 8 K, slightly below $T_c$.

\begin{figure}
		\includegraphics[width=\columnwidth]{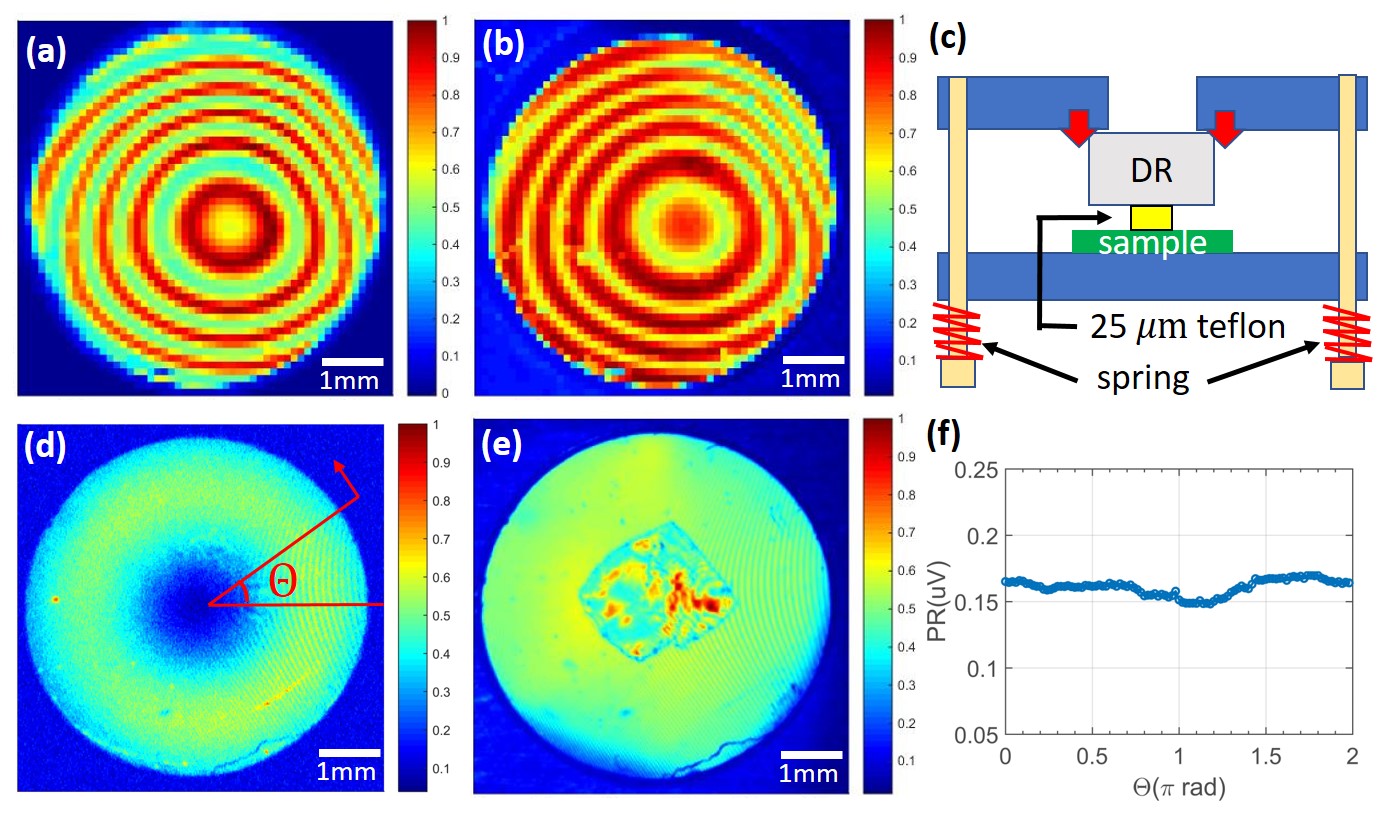}
		\caption{\label{fig:result} (a) A PR image taken with a sapphire dielectric disk placed on a Nb thin film sample at temperature $T=8.02$ K, modulation frequency $f_{mod}=10$ kHz, input microwave power $P_{in}=-10$ dBm, and TE$_{011}$ resonance frequency $f_0\sim 20.57$ GHz. The image is dominated by the Newton-ring pattern. (b) Corresponding reflectivity image, which is also dominated by a Newton-ring pattern. Note that the bright pattern in the reflectivity corresponds to the dark pattern in PR and vice versa. (c) Schematic cross section diagram (not to scale) of the teflon insertion. A 25 $\mu$m-thick teflon flake (yellow) is inserted between the sapphire disk and the sample, which makes the separation $\sim$40 times larger than the wavelength of the laser. (d) PR image taken after the teflon insertion and with a modulation frequency of 175 kHz. The Newton-ring pattern is greatly suppressed and the doughnut shape of the uniform circulating current distribution appears. (e) Corresponding reflectivity image showing the teflon and residual Newton rings. (f) Angular dependence of PR in (d) averaged over a wedge 0.02$\pi$ wide, showing $\sim$8\% anisotropy.}
\end{figure}

\par Fig.\ref{fig:result}(a) and (b) shows the image of PR and the corresponding reflectivity with a laser modulation frequency of 10 kHz. The PR is mostly isotropic but there are two features which are not predicted by the HFSS simulation of the current distribution. One is the Newton-ring pattern and the other is the absence of doughnut-shaped PR pattern. Since PR is proportional to $|\vec{j}|^2$, it is expected to have a very weak signal at the center and a strong signal at the outer radius of the sample as $|\vec{j}(x,y)|^2$ shows (Fig.\ref{fig:simulation}(c)). However, in Fig.\ref{fig:result}(a), the PR at the center is also strong.

\par For the first feature, this pattern originates from the interference of the laser light at the interface of the sapphire disk and the sample, creating Newton-rings.\cite{Hecht2002} The surface of the film and sapphire disk are smooth but not perfectly planar, creating a variable-thickness air-gap. Note that the reflectivity image has a ring pattern that is the exact complement to that of the PR image. This is expected since PR will be strong when the reflectivity is low. Therefore, to eliminate this pattern, a 25 $\mu$m-thick teflon flake is inserted and spring-loaded pressure is applied to fix it in place (see Fig.\ref{fig:result}(c)). The role of the teflon flake placed at the center of the disk is to separate the disk from the sample by a large distance compared to the wavelength of the laser light so that it suppresses the optical interference, but it is a small distance compared to the microwave wavelength so that it does not disturb the field distribution. Note that the flake is added at a location with minimal microwave current in the TE$_{011}$ mode. Also note that the spring-loaded pressure plays a crucial role to ensure reproducibility of the surface impedance measurement\cite{Mazierska1997JSuper,SYLee1997IEEE} introduced in Sec.\ref{ImpeMeausre}, which is an important complementary measurement to the PR measurement. 

\par For the second feature, the homogeneous magnitude of PR over the entire sample within the field of view of the aperture is due to the high thermal conductivity of the silicon substrate and the resulting low resolution of the PR image. Since the volume of the Nb film in this sample (300 nm thick) is small, the heat diffusion process is mainly governed by thermal properties of the silicon substrate. The thermal conductivity of silicon $\kappa_{Si}$ near $T_c$ of the Nb is of order $100$ W/mK. The thermal propagation length,\cite{Hartmann1997JAP} which is the distance that heat travels within one period of the laser intensity modulation, is $\Lambda_{Si}=\sqrt{D_{Si}/f_{mod}}=\sqrt{\kappa_{Si}/\rho c_{\rho} f_{mod}}$, where $D_{Si}$ is the thermal diffusivity, $\rho$ is the mass density, $c_{\rho}$ is the specific heat of the silicon, and $f_{mod}$ is the modulation frequency of the laser. With $f_{mod}=10$ kHz we find $\Lambda_{Si}\sim 2$ cm, which is larger than the field of view of the PR image. This large thermal propagation length of the substrate significantly reduces the resolution of the images and hence makes the magnitude of PR homogeneous throughout the sample regardless of the current distribution. To resolve this issue, a higher modulation frequency (175 kHz) is used to decrease the thermal propagation length ($\Lambda_{Si}\sim 4.7$ mm) and enhance the resolution of the PR images. Note that using a high modulation frequency decreases the signal-to-noise ratio of the PR image. Therefore, an optimal modulation frequency is determined as the lowest modulation frequency that clearly resolves the ring pattern of the circulating current as seen from Fig.\ref{fig:simulation}(c). Under such conditions it should be possible to resolve the anisotropy of the PR as well.

\par With these two modifications adopted, PR is retaken while the other conditions are fixed. As a result, the Newton-ring pattern is effectively eliminated and the doughnut-shaped PR is observed, as seen from Fig.\ref{fig:result}(d), which confirms the simulated circulating uniform current. From this Newton-ring-free high-resolution image, the angular dependence of PR is examined and shows $\sim$8\% of anisotropy (Fig.\ref{fig:result}(f)). The result is consistent with the HFSS simulated geometric anisotropy in $|\vec{j}|^2$ ($5\sim7$\% each from several mechanisms) from Sec.\ref{CurrentDensityAnisotropy}. This means that if there is any systematic anisotropy in PR larger than the $\leq 8$\% background geometric anisotropy observed, it should originate from the anisotropy of the superconducting gap function, which establishes conditions for the validity of the LSM-DR method. Note that even though there exists PR from a point defect at the $\Theta=\pi$ direction in Fig.\ref{fig:result}(d), it does not overwhelm the PR from the defect-free area and also its contribution is marginal in the PR angular dependence plot in Fig.\ref{fig:result}(f). This confirms the advantage of the new dielectric resonator method over the previous spiral resonator method, as illustrated in Fig.\ref{fig:spiralPR} 

\begin{figure}
		\includegraphics[width=\columnwidth]{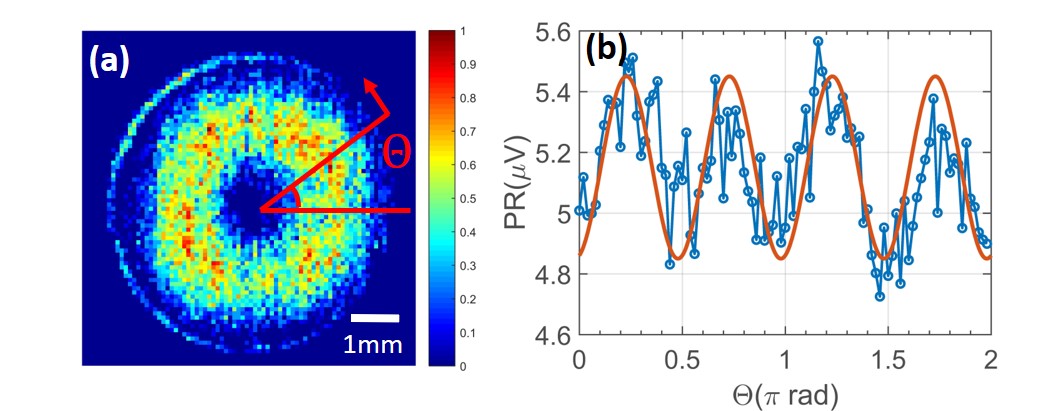}
		\caption{\label{fig:YBCOPR} (a) PR image from a 40 nm thick unpatterned YBCO thin film measured at $T=275$ mK, $P_{in}=-5$ dBm, $f_{mod}=250$ kHz, and $P_{laser} = 0.35$ mW. The definition of angle $\Theta$ is shown in red. (b) Angular dependence of PR with a $\sin^2{2\Theta}$ fit as a guide to the eye. A weak 4-fold anisotropy is visible, consistent with a 4-fold symmetry of the $d_{x^2-y^2}$ gap of YBCO.  }
\end{figure}

\par With these baseline results established, an unpatterned epitaxial and coherent $c$-axis oriented YBa$_2$Cu$_3$O$_7$ (YBCO) thin film, a representative example unconventional superconductor with a 4-fold symmetric $d_{x^2-y^2}$ gap, is examined with the dielectric resonator. The film thickness is 40 nm and it is $c$-axis normal on a sapphire substrate with CeO$_2$ buffer layer, having a coherent structure in the $a$-$b$ plane. Note that this film is heavily twinned and these boundaries can host Andreev bound states (ABS).\cite{Walter1998PRL,CRHu1994PRL,Carrington2001PRL} Note that twinning does not disrupt the 4-fold symmetry of the $d_{x^2-y^2}$ gap. As long as the $a$, $b$-directions are the same throughout the film, the 4-fold anisotropy will survive. Since the temperature dependence of the nonlinear Meissner coefficient $b_\Theta$ of a nodal superconductor becomes large as $T\rightarrow 0$,\cite{Dahm1996APL} the PR is measured at $T=275$ mK. A high modulation frequency $f_{mod}=250$ kHz is used to obtain sufficiently high resolution for the PR image. With 4 nodes in its gap function, YBCO is expected to show 4-fold symmetric PR. Indeed, as one can observe in Fig.\ref{fig:YBCOPR}(a) and (b), the PR image and its angular dependence show a 4-fold symmetric pattern with $\mathtt{\sim}$12.5\% anisotropy, making a clear contrast to the Nb case in Fig.\ref{fig:result}(d) and (f). This result confirms the ability of the LSM-PR measurement with the dielectric resonator as a gap spectroscope for unpatterned superconducting samples.

\par In the PR image, one can specifically relate a real space angle $\Theta$ to a direction in k-space. Assuming the crystallographic directions ($a$, $b$-axes) of the film follow those of the substrate, which is true for the coherent YBCO film grown on the sapphire substrate, one can deduce the crystallographic directions of the film from the cleaving directions of the substrate. With the crystallographic directions known, the $k_x$ and $k_{xy}$ directions in k-space and hence the gap nodal and anti-nodal directions can be determined. Then, the direction of the tangential current on the sample at a real space angle $\Theta$ can be matched to those k-space directions.   

\par Note that the PR along the gap nodal direction ($\Theta=0$ in Fig.\ref{fig:YBCOPR}) is smaller than that along the gap anti-nodal direction, contrary to the simple expectation from the bulk superconducting state of a $d_{x^2-y^2}$ superconductor. This is due to the paramagnetic nonlinear Meissner effect which occurs due to the ABS at the twin boundary surfaces of the YBCO film.\cite{CRHu1994PRL} This paramagnetic Meissner effect (PME) become dominant as $T\rightarrow 0$.\cite{Walter1998PRL,Carrington2001PRL} The PME gives a 45-degree rotated anisotropy in $b_\Theta(T)$ compared to that from the conventional (diamagnetic) NLME.\cite{Zhuravel2013PRL,Zhuravel2018PRB} This causes the PR image to rotate 45-degrees compared to the expectation from the conventional NLME, giving larger PR along the gap anti-nodal direction. The detailed explanation of the PR from PME can be found in Ref.\cite{Zhuravel2013PRL, Zhuravel2018PRB}

\par Regarding the comparison with the previous spiral resonator in terms of the performance as a gap nodal spectroscope, the 4-fold 12.5\% anisotropy that the dielectric resonator shows from un-patterned YBCO sample is similar to the 13\% anisotropy obtained from the patterned YBCO spiral resonator under the same measurement conditions. However, the PR from the spiral resonator shows $> 5$ times larger signal and a clearer image since the sample is self-resonant and thus very sensitive to the thermal perturbation. Thus, if a sample can be prepared in a spiral form without any defects, the spiral resonator is still preferred. However, if a sample is prone to degradation or defects under the patterning procedure, the dielectric resonator method is superior. 

\par PR contrast in the dielectric resonator method can be enhanced by utilizing several strategies. The first strategy is to increase the kinetic inductance fraction\cite{JGao2006Nucl} of the resonator (which reveals anisotropy originating from the gap) over the geometric inductance. This can be done by making the thickness of the film comparable or smaller than the magnetic penetration depth. A second strategy is to decrease the geometric factor\cite{Hein1999} which is defined by the field energy stored in the volume of the dielectric disk over that in the surface of the sample. This can be achieved by decreasing the height of the disk or measuring PR in higher TE$_{01n}$ modes with $n>1$

\subsection{Consistency check for gap nodal structure with low temperature surface impedance measurement} \label{ImpeMeausre}
\par Although examining the anisotropy of the LSM-PR is sensitive to the gap nodal properties of the superconducting samples of interest, the deduced symmetry is further confirmed by a consistency check from low temperature surface impedance measurements.\cite{Glosser1967PR,Braginsky1979IEEE,Kobayashi1991IEEE,Klein1992JSuper,Wilker1993ARFTG,Krupka1993IEEE,SYLee1997IEEE,Wingfield1997IEEE,Mazierska1997JSuper,Mazierska1998IEEE,Jacob2003IEEE,IEC61788-15,Truncik2013NatComm} In this case, the input and output microwave signals are generated and received by the Keysight vector network analyzer N5242A and the laser is turned off. The transmission lines for the microwave signal are calibrated down to the input/output ports of the dielectric resonator (p1, p2 in Fig.\ref{fig:DR}) with an in-situ cryogenic calibration technique.\cite{JenHao2013RSI} The microwave transmission signal through the dielectric resonator is represented in the form of the $S_{21}$ parameter as a function of frequency, which is defined as an transmitted output/input signal voltage ratio. By fitting this complex transmission data $S_{21}(f)$ with the phase versus frequency fitting procedure,\cite{Petersan1998JAP} one can obtain the resonance frequency $f_0$ and the quality factor $Q$. By changing the temperature of the sample and repeating this procedure, the temperature dependence of $\Delta f_0(T)=f_0(T)-f_0(T_{ref})$ and $\Delta \left(1/Q(T)\right)=1/Q(T)-1/Q(T_{ref})$ can be obtained beginning from a reference temperature $T_{ref}$. Note that the resonator is attached to the mixing chamber plate of the Bluefors XLD-400 dilution refrigerator system. The base temperature at the resonator is $\mathtt{\sim}40$ mK (with an optical window open) and the temperature is controlled up to 10 K by a resistive heater attached to the same plate where the resonator is mounted.

\par This temperature dependence of the change in the resonance frequency and quality factor can be converted to that of the surface reactance and resistance of the sample. The surface impedance gives information about the inductive superfluid response and quasi-particle dissipation of the system,\cite{Hein1999}
\begin{equation}
\Delta R_s(T) + i\Delta X_s(T) = G_{geo}\left(\Delta\left(\frac{1}{Q(T)}\right) - i\frac{2\Delta  f_0}{f_0} \right). 
\end{equation}
The temperature dependence of the surface reactance is related to the effective penetration depth $\lambda_{eff}$ of a superconducting sample by\cite{Hein1999,BBJin2002PRB,Ormeno2002PRL} 
\begin{equation}
\Delta\lambda_{eff}(T) =\frac{\Delta X_s}{\omega\mu_0}= -\frac{G_{geo}}{\pi \mu_0}\frac{\Delta f_0(T)}{f_0^2(T)}.
\end{equation}
Here, $\omega=2\pi f_0$, and $G_{geo} = \omega\mu_0\int_V dV|H(x,y,z)|^2 / \int_S dS |H(x,y)|^2$ is the sample geometric factor which can be calculated numerically using the field solution inside the resonator for each resonant mode, and was derived by Hakki and Coleman.\cite{HakkiColeman1960} The field solution assumes isotropy of the in-plane dielectric properties of the dielectric disk, which is the case for sapphire and rutile.\cite{Tobar1998JAP} In the integration, $V$ is the electromagnetic volume of the resonator and $S$ is the surface of the sample. For the sapphire resonator with the diameter $d=6.35$ mm and the height $h=3$ mm, $G_{geo}\mathtt{\sim} 391$ $\Omega$ with $f_0\mathtt{\sim}20$ GHz for the TE$_{011}$ mode. For the rutile resonator with $d=3$ mm and $h=2$ mm, $G_{geo}\mathtt{\sim}225$ $\Omega$ with $f_0\mathtt{\sim}10$ GHz for the same mode. Note that for the rutile resonator, the dielectric constant is strongly temperature dependent above 30 K so the above values for $G_{geo}$ and $f_0$ are only accurate below this temperature. A copper top plate is usually employed in the DR measurements. However, for a low $T_c$ sample ($T_c<9$ K), a Nb top plate can be used in order to increase the quality factor of the DR and thus maximize sensitivity to $\Delta R_s(T)$ and $\Delta X_s(T)$ of the sample. As long as the measurement temperature range $T<0.3T_c$ is below 1.7 K ($\approx 0.18 T_c$ of the Nb top plate), the Nb top plate shows an ignorable contribution to the temperature dependence of the surface impedance arising from the sample. Note that for the case of a sample of finite thickness, the electromagnetic fields can penetrate beyond the material and enter the substrate. This effect has been studied extensively in the past and we refer the reader to that literature.\cite{Silva1996SST,Pompeo2005SST,Pompeo2007SST}

\par From the obtained low temperature behavior of the penetration depth, one can deduce information about the low energy excitations out of the ground state, which are sensitive to the nodal structure of the gap and the cleanliness of the system.\cite{Hirschfeld1993PRB,Prozorov2006SST} Depending on the types of nodes and impurity level, $\Delta\lambda_{eff}(T)$ at low temperature (below $T/T_c<0.3$) shows a different temperature exponent $c$ when it is fitted with the $\Delta\lambda_{eff}(T)=aT^c+b$ form. Typically, a $d_{x^2-y^2}$-wave superconductor with 4 line nodes in the gap shows $c=1\mathtt{\sim}2$ depending on impurity level,\cite{Hirschfeld1993PRB} and an $s$-wave superconductor with no nodes in the gap shows $c>4$ (which is better fit to an exponential form).

\begin{widetext}

\begin{figure}
		\includegraphics[width=1\columnwidth]{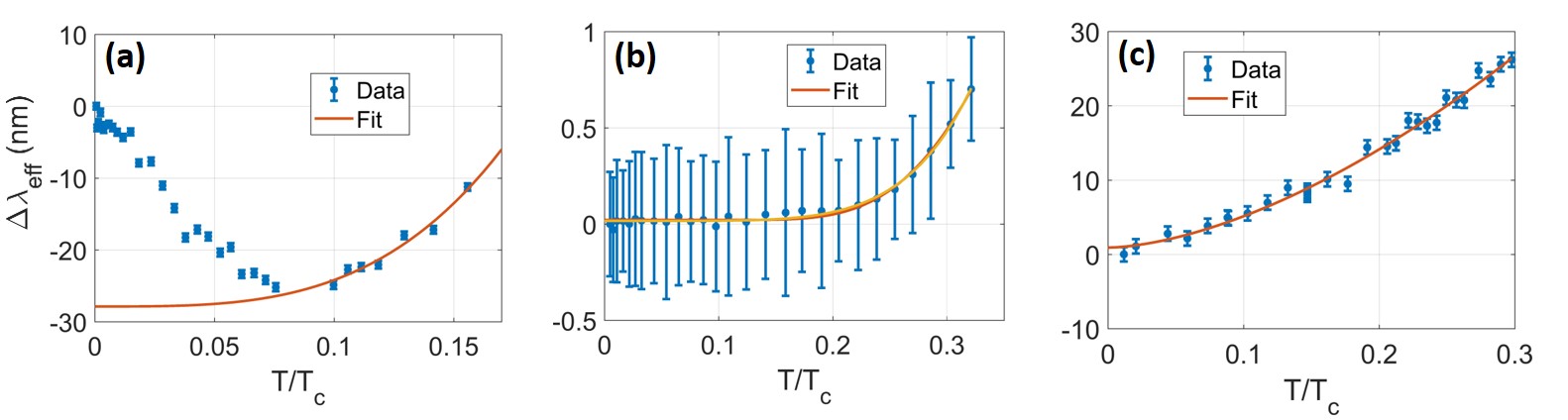}
		\caption{\label{fig:Dlambda} Low temperature behavior of the effective penetration depth $\lambda_{eff}$ for (a) the unpatterned YBCO thin film ($T_c=85$ K) measured with the sapphire resonator and Cu top plate, (b) Nb thin film ($T_c=9.25$ K) with the rutile resonator and Nb top plate (note the limited range of $\Delta \lambda_{eff}$ values), and (c) KFe$_2$As$_2$ ($T_c=3.4$ K) single crystal sample with the rutile resonator and Nb top plate. Here, the error bar in the effective penetration depth $\Delta\lambda_{eff}(T)$ is determined by the error bar of the resonance frequency $f_0(T)$. The error bar of $f_0$ is determined by a deviation which increases the rms error by 1\% when fitting is conducted to determine $f_0$. For (a), a proper $aT^c+b$ fit cannot be made due to the paramagnetic upturn of the $\Delta\lambda_{eff}$. For (b), the $aT^c+b$ fit and exponential fits give essentially the same fit line (overlapped red and orange lines). For (c), the $\Delta\lambda_{eff}$ data is fitted by $aT^c+b$ fit (red line)}
\end{figure}
\end{widetext}

\par Three example cases are shown below, first is the case of the YBCO thin film ($T_c=85$ K) complementing the PR result introduced in Sec.\ref{PR images}, which was obtained from the same film. The expectation for this well-known $d_{x^2-y^2}$-wave system is to have $c\sim 1$ in the clean limit and $c\sim 2$ in the dirty limit. As seen in Fig.\ref{fig:Dlambda}(a), above $T/T_c=0.09$, $\Delta\lambda_{eff}(T)$ shows smooth decrease as $T\rightarrow 0$ which meets the expectation for a nodal superconductor. However, below that temperature, an upturn in $\Delta\lambda_{eff}(T)$ is observed. This upturn is due to the paramagnetic Meissner effect from Andreev bound states in a $d_{x^2-y^2}$-wave superconductor which can be hosted by twin boundary surfaces.\cite{Walter1998PRL,Carrington2001PRL} YBCO thin films are known to have dense twinning boundaries that host Andreev bound states.\cite{Zhuravel2013PRL,Zhuravel2018PRB} This paramagnetic Meissner effect dominates as $T\rightarrow 0$\cite{Barash2000PRB,Zare2010PRL,Zhuravel2018PRB} and interrupts the temperature exponent study for the penetration depth by creating a non-monotonic temperature dependence.\cite{Walter1998PRL,Carrington2001PRL} This upturn in $\Delta\lambda_{eff}(T)$ is consistent with our interpretation of the PR image in Fig.\ref{fig:YBCOPR}, taken at $T/T_c=0.003$. Overall, the $\Delta\lambda_{eff}$ data is consistent with the existence of nodes in $\Delta(\vec{k})$ of YBCO.  

\par The second example is the Nb thin film used for the PR measurement in Sec.\ref{PR images}. Nb is known to have an $s$-wave pairing symmetry with a nearly isotropic gap.\cite{} Indeed, as seen from Fig.\ref{fig:Dlambda}(b), fitting with $\Delta\lambda_{eff} = aT^c+b$ below $T/T_c<0.3$ gives $c=5.73$ which is consistent with the expectation for an $s$-wave superconductor. The $\Delta\lambda_{eff}(T)$ data is also fit to the exponential temperature dependence\cite{Prozorov2006SST} $\Delta\lambda_{eff}(T)/\lambda_{eff}(0)=\sqrt{\pi\Delta(0)/2k_BT}\exp[-\Delta(0)/k_BT]$ and gives $\lambda_{eff}(0)=62.6$ nm, and $\Delta(0)/k_BT_c=1.80$. The value of $\lambda_{eff}(0)$ is similar to that of the previous literature for a 300 nm thick film ($\mathtt{\sim}75$ nm)\cite{Gubin2005PRB} and the value of $\Delta(0)/k_BT_c$ agrees well with the theoretical estimation $\Delta(0)/k_BT_c=1.78$ from BCS theory.\cite{Tinkham1996}

\par The third example is a KFe$_2$As$_2$ single crystal with $T_c=3.4$ K.\cite{YLiu2013PRB} Previous thermal and electromagnetic studies on this system indicate $d_{x^2-y^2}$-wave pairing symmetry with line nodes in the gap.\cite{Fukazawa2009JPSJ,Hashimoto2010PRB,Reid2012SST,HyunsooKim2014PRB,HyunsooKim2018RSI} Indeed, as seen from Fig.\ref{fig:Dlambda}(c), fitting $\Delta\lambda_{eff}(T)$ gives a power-law temperature exponent of $c\sim 1.6$ which is similar to a previously reported value $\mathtt{\sim}$1.4\cite{HyunsooKim2014PRB,HyunsooKim2018RSI} and consistent with line nodes in the gap.

\section{Conclusion}
\par In this work, a new type of resonator for laser scanning microscope photoresponse measurement, the dielectric resonator, is presented and validated as a new gap nodal spectroscope. First, HFSS simulation predicts that uniform circulating resonant current is generated in this setup, allowing exploration of the entire in-plane Fermi surface. The uniformity of the current density is shown to be robust against various resonator asymmetries commonly found in the experiment, demonstrating the limited extent of geometric anisotropy. An experiment with an s-wave superconducting sample shows an angular dependence of PR consistent with the expected level of geometric anisotropy. Finally, the functionality as a gap nodal spectroscope is verified with an example nodal superconductor by observing the anisotropic PR whose angular dependence is the same as that of the gap structure of the sample. Since the dielectric resonator method overcomes the defect-dominated PR issue from the previous version of the method, with its use of unpatterned samples, the new method will allow many more unconventional superconductors to be investigated for their gap nodal properties. Moreover, the ability to simultaneously measure the temperature dependence of the surface impedance of a sample brings in corroborating information about the gap nodal structure, providing a cross-confirmation with the LSM-PR result. Altogether, this method gives important information regarding gap nodal structure and hence the pairing symmetry of novel superconductors.

\begin{acknowledgments}
 We thank Lie Chen for depositing the Nb films, John Abrahams and Rahul Gogna for preliminary design and simulation of the dielectric resonator, Makariy Tanatar for preparing the mounting plate for the KFe$_2$As$_2$ crystal, and Ceraco for depositing YBCO films used in our measurement. The work of S. Bae and Y. Tan is supported by NSF Grant No. DMR-1410712 and DOE grant DE-SC0012036T. The material preparation for the Nb and YBCO films, and the measurements in the dilution refrigerator are supported by DOE grant DE-SC0018788. The contribution from A. Zhuravel is supported by Volkswagen Foundation grant No. 90284.
\end{acknowledgments}

\bibliography{Ref_DR_RSI_v2}

\end{document}